\documentclass[11pt]{article}
\usepackage{graphicx} 
\usepackage{lineno}
\usepackage{amsmath,amsfonts,amssymb}
\usepackage[left = 3.5cm, right= 3.5cm, top = 3.5cm, bottom = 3.5cm]{geometry}
\usepackage[authoryear,round]{natbib}
\usepackage{xcolor}
\usepackage{tikz}
\usepackage{tikz-cd}
\usepackage{float}
\usepackage{lscape}
\usepackage{bm}
\usepackage{hyperref}
\title{A Continuous-Time Markov Chain Framework for Population Size Estimation from Multi-List Data: Accounting for Absorbing Lists and Asymmetric Interactions}
\author{Oph\'elie Schaller$^{1,*}$, Andrew Titman$^1$, Rachel McCrea$^1$}

\begin{document}

\begin{titlepage}
    \maketitle
           
\begin{itemize}
    \item[] $^{1}$School of Mathematical Sciences, Lancaster University, United Kingdom
\end{itemize}

$^{*}$Corresponding author: o.schaller@lancaster.ac.uk \\
       
\begin{center}
    \subsubsection*{Abstract} 
\end{center} 
We introduce a continuous-time Markov chain framework for estimating population size from multi-list data, which allows directional interactions to be modelled and can accommodate absorbing lists, such as death records, or more general data collection processes. 
The standard model of the continuous-time Markov chain framework and the log-linear model for multi-list data are equivalent when lists are independent and we show empirically that they give similar results in the presence of dependencies between lists. 
Through a simulation study, we highlight the need to account for an absorbing list by using the Markov model or the log-linear model with forced absorbing interactions, observing biased estimates of the population size otherwise. 
We motivate our approach with an epidemiological dataset concerning individuals suffering from a first ever stroke in North-West England, in which one of the lists is a death record. We illustrate a further use of our approach considering a case of ordered lists on drug use data from the City of London.
 \\

\noindent \textbf{Keywords}: hidden population; multiple systems estimation; population size; continuous-time Markov chain; absorbing list. 
\end{titlepage}

\section{Introduction} \label{introduction}

Estimating the size of hard-to-reach populations is of high interest to inform public policy \citep{bird2018multiple}. When multiple partial observations of a population are combined through a record linkage procedure, multiple systems estimation (MSE) or capture-recapture methods can be used to estimate the number of hidden individuals. While capture-recapture tends to be applied to wildlife populations, MSE has been used in epidemiology \citep{hook1995apptoepidemiology}, on populations of drug users \citep{king_bird_overstall, djennadjrsssa2024}, victims of modern slavery or human trafficking \citep{bales2015modern}, casualties in a war \citep{Kosovo2002} or homeless individuals admitted to the hospital \citep{homelessness_hospital}. 
Both approaches originated from the same models, however the literature diverged to account for different data collection types. 

In an ecological context, the partial observations tend to be of the same nature, conducted repeatedly at precise points in time, originally in the form of captures, release, and recaptures. 
In this case, models have been developed for open populations to account for death, emigration and birth \citep{jolly1965, seber1965, Cormack1989, worthington2019}  and to include temporal, behavioural or individual heterogeneity \citep{otis1978}. 
 Note that in some situations human data takes a form similar to ecological data and can be analyzed using a capture-recapture analysis \citep{brechtwickens1993sagejournal}.
For a review of capture-recapture techniques, see for example \citet{KingMcCrea2019}.

Most often, within a social context, data sources are administrative lists from independent organisations collecting data in no known order over the same period of time. The resulting accessible dataset can be displayed in an incomplete $2^k$ contingency table with $k$ being the number of lists, where each binary category represents whether individuals have been recorded on a corresponding list at some unknown point during the recording period. The value of one table cell is missing: individuals observed nowhere. 
Most often, a log-linear model is fitted to multi-list data \citep{Fienberg1972} and values of the cells of the contingency table are assumed to result from a Poisson distribution \citep{SandlandCormack1984}. When using this model, it is assumed that the target population is closed, that the probability of being recorded on a list is homogeneous over individuals, that individuals are perfectly matched between lists and that there is no highest order interaction between lists. 

In their review of applications of MSE in epidemiology, \citet{hook1995apptoepidemiology} state the closure assumption as one of its main limitations.  \citet{worthington2021} encouraged further development of MSE through a comparison with ecological capture-recapture methods, highlighting the necessity to model open populations as one of the main areas of MSE which could be improved. 
To our knowledge, open population models have not yet been proposed within the MSE context and the primary reason for this is the lack of temporal information in the data.

In some cases, the target population is known to be at least partially open, which clearly violates the closure assumption. This is the case when one of the lists is absorbing, i.e. the list records individuals as they leave the population, such as an emigration or death record. We will show in Section \ref{section_simulation} that in that case not accounting for an absorbing list will lead to an overestimation of the total population size. Additionally, this situation does not respect the homogeneity assumption, as not all individuals stay observable for the same amount of time. Death records are for example frequent in epidemiological applications (e.g., \citealt{hook2000underestimation, barocas2018estimated, djennadjrsssa2024}). Motivated in particular by a dataset on individuals suffering from their first ever stroke from \cite{hook2000underestimation}, which we introduce below, we propose a new framework for MSE. 
Even though we only have access to data at one point in time, we model the continuous process of the collection of data using Markov chains. This allows us to take into account the particular nature of absorbing lists, as well as to more generally model effects in the collection of data. We give an additional example on data on drug use in the City of London where some of the lists are ordered over time. 

In Section \ref{section_markovmodel}, we describe the continuous-time Markov chain model for multi-list data, demonstrating how the framework can be used to account for an absorbing list and its adaptability to other scenarios. In Section \ref{section_simulation}, we present a simulation study comparing estimates obtained between the proposed Markov model and the standard log-linear model on multi-list data, both in a general case and in the presence of an absorbing list. In Section \ref{section_applications}, we present results of our analysis on the datasets introduced below. Finally, we discuss results and further research avenues for this new Markov framework in Section \ref{section_conclusion}.

\subsubsection*{First ever stroke in Northwest England 1994-1995} \label{northwest_england_data_subsection}

        Our main motivating dataset concerns individuals suffering from their first ever stroke in Northwest England between July 1994 and June 1995 \citep{du1997} and is displayed in Table \ref{firsteverstroke_data}. This particular example has been cross-classified from five lists: general practice reports; hospital records; death registers; other groups such as rehabilitation services, nursing homes and support services. \citet{du1997} only take into consideration cases recorded in at least one of those lists to estimate the total number of strokes in the designated practices. In \citet{hook2000underestimation}, the data is used for an internal validity study where a MSE log-linear model is used on a subset of those lists and compared with the known value. This analysis does not consider the absorbing properties of the death list, which as we will show in Section \ref{section_simulation}, could lead to an underestimation of the population size.

		\begin{table}
			\begin{center}
				\begin{tabular}{c c c c c | c }
					\hline
					$L_1$ & $L_2$ & $L_3$ & $L_4$ & $L_5$ & counts \\
					\hline
					0 & 0 & 0 & 0 & 0 & $?$\\
					1 & 0 & 0 & 0 & 0 & 158\\
					0 & 1 & 0 & 0 & 0 & 68\\
					0 & 0 & 1 & 0 & 0 & 65\\
					0 & 0 & 0 & 1 & 0 & 25\\
					0 & 0 & 0 & 0 & 1 & 15\\
					1 & 1 & 0 & 0 & 0 & 44\\
					1 & 0 & 1 & 0 & 0 & 26\\
					1 & 0 & 0 & 1 & 0 & 20\\
					1 & 0 & 0 & 0 & 1 & 6\\
					0 & 1 & 1 & 0 & 0 & 69\\
					0 & 1 & 0 & 1 & 0 & 8\\
					0 & 1 & 0 & 0 & 1 & 11\\
					\hline
				\end{tabular}
				\quad
				\begin{tabular}{c c c c c | c }
					\hline
					$L_1$ & $L_2$ & $L_3$ & $L_4$ & $L_5$ & counts \\
					\hline
                    0 & 0 & 1 & 1 & 0 & 3\\
					0 & 0 & 1 & 0 & 1 & 1\\
					1 & 1 & 1 & 0 & 0 & 39\\
					1 & 1 & 0 & 1 & 0 & 43\\
					1 & 1 & 0 & 0 & 1 & 6\\
					1 & 0 & 1 & 1 & 0 & 1\\
					1 & 0 & 1 & 0 & 1 & 2\\
					1 & 0 & 0 & 1 & 1 & 1\\
					0 & 1 & 1 & 1 & 0 & 12\\
					0 & 1 & 1 & 0 & 1 & 5\\
					0 & 0 & 1 & 1 & 1 & 1\\
					1 & 1 & 1 & 1 & 0 & 11\\
					1 & 1 & 1 & 0 & 1 & 2\\
	
					\hline
				\end{tabular}
			\end{center}
            		\caption[Data on strokes in Northwest England.]{Counts of cases of first ever stroke recorded in Northwest England between July 1994 to June 1995. $L_1$: general practices reports; $L_2$: statutory hospital activity analysis minimum data set; $L_3$: death certificates; $L_4$: hospital wards registers and discharge data set; $L_5$: other sources.}
			\label{firsteverstroke_data}
		\end{table}
        
\subsubsection*{Drug use in the City of London, 2018-2019}

        The second dataset we will analyse is formed of two administrative sources for opiate and/or crack cocaine use (OCU) in the City of London for the financial year 2018-2019 \citep{djennadjrsssa2024} and is displayed in Table \ref{drugabuse_data_cityoflondon}.
        The lists correspond to community treatment and the criminal justice system. The first list contains a record of both individuals staying in community treatment until the end of the year and individuals leaving community treatment. 
        As this was previously analysed as one list with a covariate parameter (staying in or leaving treatment), we propose an alternative  continuous-time model for these data by considering in treatment and out of treatment as two different states. This case study serves as an illustration of the adaptability of the proposed Markov framework to better account for the dynamics underpinning the data collection processes. 

        \begin{table}[H]
			\begin{center}
				\begin{tabular}{c | c c c c c c c c}
					\hline
					$L_1$ & 0 & 1 & 0&0&1&1&0&1\\
					$L_2$ & 0 &  0 & 1&0&1&0&1&1\\								
					$L_3$ & 0 & 0 & 0&1&0&1&1&1\\
					
					\hline 
					& ? & 154 & 7 & 0 & 2 & 0 & 5 & 9  \\
					\hline 
				\end{tabular}
			\end{center}
            \caption[Data on drug use in the City of London.]{Number of individuals aged 15-64 in the City of London recorded for opiate and/or crack cocaine use in years 2018/2019 with lists $L_1,L_2,L_3$ corresponding respectively to criminal justice system, took part in community treatment and left community treatment. }
			\label{drugabuse_data_cityoflondon}
		\end{table}

\section{Continuous-time Markov chain framework for multiple systems estimation} \label{section_markovmodel}

In Section \ref{subsection_markovmodel}, we introduce a continuous-time Markov chain framework for population size estimation from multi-list data. It is mathematically equivalent to the log-linear framework introduced by \cite{Fienberg1972} when lists are independent, and both frameworks give close results when interactions are present (see Appendix \ref{markov_loglin_equiv} and Section \ref{subsection_markovvslogline}). Furthermore, we show that the Markov framework can be easily adapted to account for absorbing lists in Section \ref{subsection_account_abso_list} and to more general asymmetric interactions and other data collection processes in Section \ref{subsection_other_uses}.

\subsection{General continuous-time Markov chain model for MSE} \label{subsection_markovmodel}

We define the following continuous-time Markov chain model for multi-list data. Our primary parameter of interest is the unknown population size, which we denote by $N$. For data from $k$ lists $L_1, L_2, \dots, L_k$, we choose the state-space $\mathcal{L} := \mathcal{P}(\{L_1, L_2,...,L_k \})$, the power set of the set of lists. 
Let $\mathcal{I} := \mathcal{L} \setminus \{\emptyset\}$. For $I \in \mathcal{I}$ we denote by $n_I$ the observed count of individuals appearing in all and only lists in $I$. 
We suppose that the data on the target population has been collected from time $t_0$ until time $t_f$. Without loss of generality, we assume that $t_0 = 0$ and $t_f = 1$. We say that an individual belongs to a state $I \in \mathcal{L}$ at time $t \in [t_0, t_f]$ if that individual has been recorded in all and only the lists in $I$ at time $t$. We assume that all individuals are in the state $I_0 := \emptyset$ at time $t_0$, i.e., before any collection of the data, all individuals have not been observed by any list. It is supposed that all the individuals of the target population have the same probability $p_{t,I}$ of being in state $I$ at time $t$.
We suppose that there exists a continuous random process $(X_t)_{t \geq 0}$ such that 
$$\mathbb{P}( X_t = I) = p_{t,I}.$$

\begin{figure} 
		
		\[\begin{tikzcd}
			&&& {\boxed{L_1}} &&& {\boxed{L_1,L_2}} \\
			\\
			{\boxed\emptyset} &&& {\boxed{L_2}} &&& {\boxed{L_1,L_3}} &&& {\boxed{L_1,L_2,L_3}} \\
			\\
			&&& {\boxed{L_3}} &&& {\boxed{L_2,L_3}}
			\arrow["{\lambda_2\mu_{12}}"{description}, from=1-4, to=1-7]
			\arrow["{\lambda_3\mu_{13}}"{description, pos=0.3}, shift right, from=1-4, to=3-7]
			\arrow["{\lambda_3\mu_{13}\mu_{23}\mu_{123}}"{description, pos=0.4}, from=1-7, to=3-10]
			\arrow["{\lambda_1}"{description}, from=3-1, to=1-4]
			\arrow["{\lambda_2}"{description}, from=3-1, to=3-4]
			\arrow["{\lambda_3}"{description}, from=3-1, to=5-4]
			\arrow["{\lambda_1\mu_{12}}"{description, pos=0.7}, from=3-4, to=1-7]
			\arrow["{\lambda_3\mu_{23}}"{description, pos=0.3}, from=3-4, to=5-7]
			\arrow["{\lambda_2\mu_{12}\mu_{23}\mu_{123}}"{description}, from=3-7, to=3-10]
			\arrow["{\lambda_1\mu_{13}}"{description, pos=0.8}, from=5-4, to=3-7]
			\arrow["{\lambda_2\mu_{23}}"{description}, from=5-4, to=5-7]
			\arrow["{\lambda_1\mu_{12}\mu_{13}\mu_{123}}"{description, pos=0.4}, from=5-7, to=3-10]
		\end{tikzcd}\]
		\caption[Graph of Markov process for three lists.]{Graph of transition between states in $\mathcal{L} = \{L_1, L_2, L_3\}$.} \label{mc_full_graph}
	\end{figure}
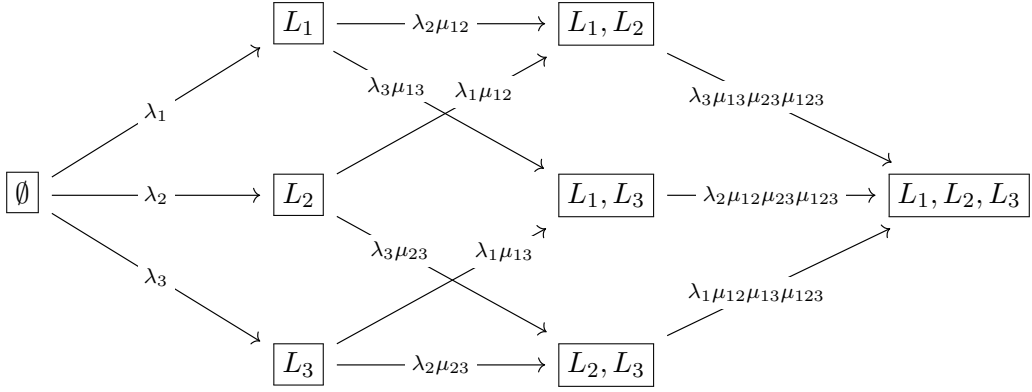

Furthermore, we assume that $(X_t)_{t \geq 0}$ is Markov$(\phi_0, Q)$, a continuous-time Markov chain with initial distribution $\phi_0$ and generator matrix $Q$, where $\phi_0(\emptyset) = 1$ and $\phi_0 (I) = 0$ for any $I \in \mathcal{I}$ and where $Q = (q_{IJ} : I,J \in \mathcal{L})$ has the following restriction: $q_{IJ} = 0$ if $I \not\subseteq J$ or if $|J| - |I| > 1$.
More precisely, we propose the following parameterisation of the matrix $Q$. Let $I,J \in \mathcal{L}$ be two subsets of lists such that $I = \{L_{i_1}, L_{i_2}, \dots , L_{i_l} \}$ and $J = \{L_{i_1}, L_{i_2}, $..., $L_{i_l}, L_{i_{l+1}} \}$ for $i_1, \dots , i_{l+1} \in \{1,\dots, k\}$. Let $\lambda_1, \lambda_2, \dots , \lambda_k \in [0, \infty[$ denote transition parameters corresponding to lists $L_1, \dots, L_k$ respectively an interaction parameters $\mu_{K} \in [0, \infty[$ for $K \in \mathcal{I}$ with $|K| \geq 2$. We define 
$$q_{IJ} := \lambda_{i_{l+1}} \prod_{\substack{K \in \mathcal{I}, K \subseteq J \\ |K| \geq 2, L_{i_{l+1}} \in K }} \mu_K.$$
For any $I \neq J$ not satisfying the conditions above, we set $q_{IJ} := 0$. For the case $I = J$, we set $q_{II} := - \sum_{J \in \mathcal{L} \setminus \{I\} } q_{IJ}$. In order to simplify notation, we will write $\mu_{i_1, i_2, \dots, i_l}$ for $\mu_{\{L_{i_1}, L_{i_2}, \dots, L_{i_l}\}}.$ 
In the case $k = 3$, the corresponding Markov graph is displayed in Figure \ref{mc_full_graph}. 

    \subsubsection*{Fitting the continuous-time Markov model to multi-list data}
	As we are interested in data which does not contain any information on intermediate counts at a time $t \in ]t_0,t_f[$, we only want to model the probabilities $p_{t_f,I}$ for $I \in \mathcal{L}$ of being in a state at the end of the recording period. Let $P = (p_{I,J} : I,J \in \mathcal{L})$ be the solution of the differential matrix equation
    \begin{equation} \label{diff_mat_equ}
        P'(t) = P(t)Q, \; P(0) = I_{k \times k}.
    \end{equation}
    The solution of Equation \ref{diff_mat_equ} is given by the exponential matrix $e^{tQ}$ (see e.g. \citealt{norris97}). Recall that if $V$ is the matrix of eigenvectors of $Q$ and $D$ the diagonal matrix of eigenvalues of $Q$ such that $Q = VDV^{-1}$, one has that $e^{tQ} = Ve^{tD}V^{-1}$ (see e.g. \citealt{norris97}). 
	By properties of continuous-time Markov chains (see e.g. \citealt{norris97}) and as we assumed all individuals are not seen anywhere at time $t_0 = 0$, we have 
	$$\mathbb{P}(X_{t_f} = I) = \mathbb{P}(X_{t_f} = I | X_0 = \emptyset) = p_{\emptyset,I}(t_f).$$
	As $t_f$ is set to be equal to $1$, the first row of the matrix $P(1)$ models the vector of probabilities for an individual in the population of being in each cell of the contingency table. 
Assuming each of the $N$ subjects in the population independently follow the Markov chain model, we have that
$$(n_I)_{I \in \mathcal{L}} \mid N \sim \mbox{Multinomial}\{N, (p_{\emptyset, I}(1))_{I \in \mathcal{L}})\}.$$ 
We let $f(N,(p_{\emptyset,I}(1))_{I \in \mathcal{I}}) = f(N,(p_{\emptyset,I}(1))_{I \in \mathcal{I}}| (n_I)_{I \in \mathcal{I}})$ denote this multinomial likelihood, $p_{obs} = \sum_{I \in \mathcal{I}} p_{\emptyset,I}(1)$ denote the probability of being observed and $n = \sum_{I \in \mathcal{I}} n_I$ denote the total number of observed individuals. We can separate the multinomial likelihood into two components as follows 
$$f(N,(p_{\emptyset,I}(1))_{I \in \mathcal{I}}) = f_1(N,  p_{obs}) f_2((p_{\emptyset,I}(1))_{I \in \mathcal{I}}), $$
with
$$f_1(N,p_{obs}) = \frac{N!}{n!(N-n)!} p_{obs}^n (1-p_{obs})^{N-n}$$
a binomial term and with 
$$f_2((p_{\emptyset,I}(1))_{I \in \mathcal{I}}) = \frac{n!}{\prod_{I \in \mathcal{I}}n_I!}\prod_{I \in \mathcal{I}} \left( \frac{p_{\emptyset,I}(1)}{p_{obs}} \right)^{n_I}$$
a multinomial term independent of $N$. 
Estimation of $N$ can then proceed either by directly maximising the multinomial likelihood $f$, including $N \geq \sum_{I \in \mathcal{I}} n_I$, as an additional parameter, or by first maximising the term $f_2$ to obtain conditional estimates of the Markov parameters and hence probabilities $(\hat{p}^C_{\emptyset,I}(1))_{I \in \mathcal{I}}$, to then maximise $f_1$ given those estimates. The conditional maximal likelihood estimate of $N$ will in this case be given by 
$$\hat{N}^C = \frac{n}{\hat{p}^C_{obs}}.$$
Both multinomial methods are asymptotically equivalent \citep{cormackjupp1991}. Alternatively, we can treat the observed counts $(n_I)_{I \in \mathcal{I}}$ as independent Poisson counts with rates $N p_{\emptyset, I}(1)$ for $I \in \mathcal{I}$, which is equivalent to using the conditional multinomial likelihood \citep{cormackjupp1991,SandlandCormack1984}. 
The Poisson approach leads to log-likelihood and AIC values that are directly comparable with those obtained through the log-linear model and we will hence prefer this method in our simulation studies and in the analysis of the data presented in this paper.

          Note that the Markov model is equivalent to the log-linear model when the lists are independent (see Appendix \ref{markov_loglin_equiv} for details). 
When there is dependence between lists, we observe empirically through a simulation study that both approaches most often select the same non-neutral interaction parameters and that the selected models give similar AIC values and lead to similar estimates of the total population size (see Section \ref{section_simulation}).

    Several models can be considered by the choice of which interaction parameters will not be equal to 1. 
	In line with the approach for the log-linear case, we restrict the set of considered models to those satisfying the hierarchical principle. To ensure identifiability of the parameters we will always constrain the full interaction parameter $\mu_{12 \dots k}$ to be equal to 1, allowing for a maximal number of non-neutral parameters of $2^{k} - 1$ in the case of the saturated model.

    	Several assumptions are made under this framework. As with the log-linear model, one assumes each individual in the population has the same probability of being in each of the states. In this case, more precisely, we suppose each individual has the same probability $p_{\emptyset,I}(t)$ of being in any state $I$ at any time $t \in [t_0, t_f]$. Furthermore, the continuous-time Markov chain assumption implies that for some operational time scale, the capture times on that scale are exponentially distributed. 
        
     The main limitation of the method lies in its computational cost, which is much higher than that of the log-linear model as it requires a matrix inversion with matrix size $2^k \times 2^k$ where $k$ is the number of lists. This implies that the use of a bootstrap procedure will not be easily accessible especially when the number of lists gets larger.

	The new possibilities that arise from this approach can be highlighted by the ability of the model to describe data collection processes and to create asymmetric interactions between lists. In Section \ref{subsection_account_abso_list}, we present a modification of this model which enables an absorbing list to be explicitly accounted for, whilst in Section \ref{subsection_other_uses}, we explore further possibilities for the Markov model, in particular the case of ordered lists.

\subsection{Accounting for an absorbing list} \label{subsection_account_abso_list}

We now use the continuous-time Markov chain framework to account for the presence of an absorbing list in multi-list data. In this setting, for data from $k$ sources $L_1, \dots, L_k$, the fact that source $L_i$, for some $1 \leq i \leq k$, is absorbing can be defined by the impossibility for an individual to move to any other state once they are in a state containing list $L_i$. Modelling an absorbing list therefore requires some directional interaction parameters to be constrained to zero. In practice this requires setting some rows of the generator matrix $Q$ defined in Section \ref{subsection_markovmodel} to zero, and hence this removes the corresponding arrows of Figure \ref{mc_full_graph}. 

Explicitly, we construct $Q^i = (q_{IJ}^i : I,J \in \mathcal{L})$, the $Q$-matrix corresponding to a scenario with $k$ lists $L_1, \dots, L_k$ such that the list $L_i$ is absorbing, in terms of parameters $\lambda_1, \dots, \lambda_k$ and $(\mu_K)_{K \in \mathcal{L}, |K] \geq2}$ as follows. Let $Q = (q_{IJ} : I,J \in \mathcal{L})$ be the $Q$-matrix corresponding to standard multi-list data scenario with $k$ lists as constructed in Section \ref{subsection_markovmodel}. For $I, J \in \mathcal{L}$, we define 

	\begin{align*}
		q^i_{IJ} := \begin{cases}
			q_{IJ}, & \text{if } L_i \notin I; \\
			0, & \text{if } L_i \in I .
		\end{cases} 
	\end{align*}

\begin{figure}[H]
			\begin{center}
				\[\begin{tikzcd}
					&&& {\boxed{L_1}} &&& {\boxed{L_1,\mathbf{L_2}}} \\
					\\
					{\boxed\emptyset} &&& {\boxed{\mathbf{L_2}}} &&& {\boxed{L_1,L_3}} &&& {\boxed{L_1,\mathbf{L_2},L_3}} \\
					\\
					&&& {\boxed{L_3}} &&& {\boxed{\mathbf{L_2},L_3}}
					\arrow["{\lambda_2\mu_{12}}"{description}, from=1-4, to=1-7]
					\arrow["{\lambda_3\mu_{13}}"{description, pos=0.3}, shift right, from=1-4, to=3-7]
					\arrow["{\lambda_1}"{description}, from=3-1, to=1-4]
					\arrow["{\lambda_2}"{description}, from=3-1, to=3-4]
					\arrow["{\lambda_3}"{description}, from=3-1, to=5-4]
					\arrow["{\lambda_2\mu_{12}\mu_{23}\mu_{123}}"{description}, from=3-7, to=3-10]
					\arrow["{\lambda_1\mu_{13}}"{description, pos=0.8}, from=5-4, to=3-7]
					\arrow["{\lambda_2\mu_{23}}"{description}, from=5-4, to=5-7]
				\end{tikzcd}\]
			\end{center}
             \caption[Graph of Markov process for an absorbing list.]{Graph of transitions between sates in $\mathcal{L} = \{L_1, L_2, L_3 \}$ with absorbing list $L_2$.} \label{absorbing_graph}
		\end{figure}
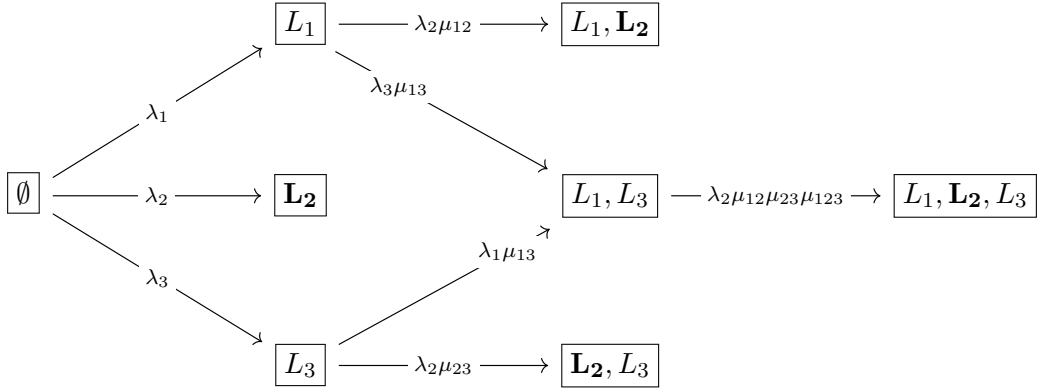

\noindent In the case where $k = 3$ and $L_2$ is absorbing, we have the Markov graph displayed in Figure \ref{absorbing_graph}.

In this setting, we no longer assume that the population is fully closed. However, some assumptions on the closure of the population still hold. We still have to assume that no observed individual enters or leaves the population of interest without being seen to do so. In most cases, this is clearly not necessarily true. The minimum assumptions on closure of the population would be that no individual who is not in state $\emptyset$ leaves the population without being seen, and that the rate of new individuals secretly entering and leaving the population is the same. In other words, that the number of individuals in each state does not change. 

In Section \ref{section_simulation}, we show in a simulation study that failing to account for an absorbing list when using MSE can result in biased estimates of the population size. We illustrate our approach with a real life epidemiological example in Section \ref{section_applications}.

\subsection{Further uses of the Markov model for multiple systems estimation} \label{subsection_other_uses}

\subsubsection*{Accounting for ordered data collection} \label{subsection_account_ordered}

\begin{figure} [H]
			
			\[\begin{tikzcd}
				&&& {\boxed{L_1}} &&& {\boxed{L_1,L_2}} \\
				\\
				{\boxed\emptyset} &&& {\boxed{L_2}} &&& {\boxed{L_2,L_3}} &&& {\boxed{L_1,L_2,L_3}}
				\arrow["{{\lambda_2\mu_{12}}}"{description}, from=1-4, to=1-7]
				\arrow["{{\lambda_3\mu_{13}}}"{description, pos=0.4}, from=1-7, to=3-10]
				\arrow["{{\lambda_1}}"{description}, from=3-1, to=1-4]
				\arrow["{{\lambda_2}}"{description}, from=3-1, to=3-4]
				\arrow["{{\lambda_1\mu_{12}}}"{description}, from=3-4, to=1-7]
				\arrow["{{\lambda_3}}"{description}, shift right, from=3-4, to=3-7]
				\arrow["{{\lambda_1\mu_{13}\mu_{12}}}"{description}, from=3-7, to=3-10]
			\end{tikzcd}\]
            \caption{Graph of transition between states in $\mathcal{L} = \{L_1, L_2, L_3 \}$ ordered lists $L_2$ and $L_3$.}\label{drugabuse_graph1}
		\end{figure}
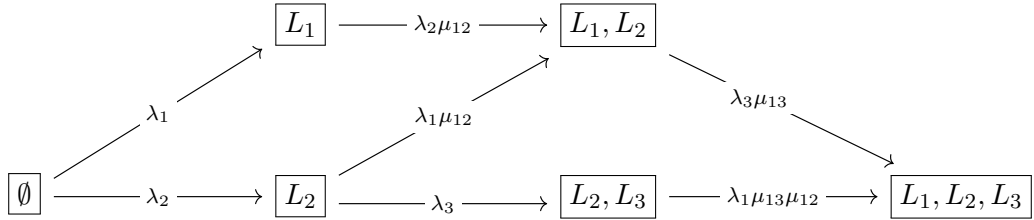

    The continuous-time Markov framework for MSE is highly flexible and can be adapted to further scenarios. 
For example, we now show how we can account for a multi-list scenario in which some of the lists are ordered over time. Suppose that we have $k$ data sources $L_1, \dots, L_k$ such that $L_i$ and $L_j$ are ordered, i.e. such that no individual can be recorded on $L_j$ without having been previously recorded on list $L_i$. Once again, we build a continuous-time Markov chain describing the recording process of the total population by modifying the construction given in Section \ref{subsection_markovmodel}. This time, we reduce the state-space to $\mathcal{L}_{i\rightarrow j} :=  \mathcal{L} \setminus \{ I \}_{L_j \in I, L_i \notin I}$. We define the generator matrix $Q_{i \rightarrow j} = (q_{IJ}: I,J \in \mathcal{L}_{i \rightarrow j})$ to be the submatrix of $Q$ from Section \ref{subsection_markovmodel} for pairs $I,J \in \mathcal{L}_{i \rightarrow j}$. In this case, some interaction parameters will be redundant and can be omitted. More precisely, $\lambda_j$ and $\mu_{ij}$ will always arise in the same terms of the matrix $Q_{i \rightarrow j}$, hence the interaction term $\mu_{ij}$ can be omitted. Similarly, the terms $\mu_{i_1 \dots i_l i j}$ and $\mu_{i_1 \dots i_l j}$ will always appear together and only the lowest order term can be considered non-neutral. We show in Figure \ref{drugabuse_graph1} an example of a Markov graph for a scenario with three lists $L_1, L_2, L_3$ such that $L_2$ and $L_3$ are ordered over time.

Similarly as for the log-linear case or as for the standard continuous-time Markov chain framework for MSE, the total number of parameters exceed by one the number of data points (i.e. the number of non-empty states). Indeed, the total number of states in $\mathcal{L}_{i \rightarrow j}$ is equal to $2^k - 2^{k-2}$, which is equal to the total number of non-redundant parameters. For this reason, one parameter at least amongst the highest order interaction terms has to be equal to 1. In Section \ref{section_applications}, we present the results of the analysis of three-list data on drug users in the City of London in which two lists are ordered.

\subsubsection*{General asymmetric interactions and further}

    The continuous-time Markov chain model can be used for further more general scenarios. 
    For example, when the number of lists is high enough and with the necessary restrictions to not include some of the higher-order interactions, we can estimate two different orientated interactions between lists. Instead of being limited to one interaction $\mu_{12}$ between lists $L_1$ and $L_2$, one could now estimate both $\mu_{\overset{\rightarrow}{12}}$ and $\mu_{\overset{\rightarrow}{21}}$. 
	This is interesting in cases where strong asymmetry is suspected but where none of the directional interaction is automatically zero or one, contrarily to the cases of absorbing or ordered lists presented above, for example in the presence of referrals between lists \citep{jonesreferral2014}.

    This framework also allows us to be flexible with the time periods in which the collection from different data sources took place.  For example, \citet{barocas2018estimated} work with a several lists with observation period between January 1, 2011 and December 31, 2015 and with a list with observation period being between January 1, 2011 and September 30, 2015. This time period difference can be modelled within the Markov chain model framework. 

    We think that further open population models could be explored, for example by adding an unobserved state for individuals who have left the population. In this case, if we consider that any individual in any state has the same probability of leaving the population in a similar timeline, we could model this situation with one additional transition parameter, which is feasible when the number of lists is high enough. Similarly, we could explore the possibility of modelling people entering the population by adding a non-observable state with transition possibility to the observable but non-observed state.  

\section{Simulation study} \label{section_simulation}

We present results from two simulation studies. In Section \ref{subsection_markovvslogline}, we empirically compare the general continuous-time Markov chain model and the standard log-linear model for MSE in the presence of dependencies between lists. In Section \ref{subsection_simabso}, we compare total population size estimates of simulated multi-list data with an absorbing list obtained by using the naive standard log-linear model, the continuous-time Markov chain model for an absorbing list and the log-linear model with forced interactions.

\subsection{Continuous-time Markov chain model versus log-linear model} \label{subsection_markovvslogline}

We compare results obtained from the continuous-time Markov chain model and the standard log-linear model when applied to the same simulated data. We focus our comparison on the non-neutral interaction parameters selected, the AIC values obtained by the chosen models and the total population sizes estimates. We simulated the data from the Markov model and considered four-list scenarios, varying the number of two-way interaction parameters present in the generating model between 1 and 6. 
We considered two stepwise forward model selection approaches: (i) selecting the more complex model if its AIC score is at least two points smaller than the AIC score of the base model and if its score is the best amongst candidate models - this difference in AIC is often preferred as a difference of less than 2 does not consist substantial evidence that the model with the lowest AIC is better at describing the data \citep{burnham2002model} ; and (ii) an `accelerated' forward selection where the more complex model is chosen if its AIC score is simply smaller than that of the base model. 

Table \ref{diff_estimates_ms_acc} displays results from the case presenting the biggest differences between the methods. For additional results and the details of the scenarios parameters, see Tables \ref{mc_scenario_unique} to \ref{diff_AIC_forward} and Figures \ref{comparison_figure4} and \ref{comparison_figure5} in Appendix \ref{appendix_comparison_sim}. In general, we observed that results obtained from both methods are close. Results are the closest when there is no model selection and closer when the model selection contains fewer steps. Furthermore, results are closer when the number of interactions contained in the generator model is smaller. In all the scenarios considered, the same models were selected more than $90 \%$ of the time. We conclude that the outcomes of the Markov model and log-linear are similar when model assumptions are respected. 

	\begin{table} [tb!]
		\begin{center}
			\begin{tabular}{| l |  c | c c c c   | }
				\hline 
				$\#$int. & $N$ & med. $\Delta \hat{N}$ &  $[1_{st} Q.,3_{rd} Q.] \Delta \hat{N} $ &  $Q_{0.95} |\Delta\hat{N}|$   & Diff.mo.\\
				\hline
				1 & 500 & 0.001 & [-0.607, 0.303]  & 11.43   & 21/500 \\
				2 & 500 & 0.002 & [-0.468, 0.443]  & 11.19   & 24/500\\
				3 & 500 & -0.025  & [-0.444, 0.304]  & 5.018   & 15/500 \\
				4 & 500 & -0.016  & [-0.810, 0.292]  & 19.070   & 22/500 \\
				5 & 500 & -0.242  & [-1.989, 0.162]  & 19.666   & 31/500 \\
				6  & 500 & -0.817 & [-4.097, 0.490]  & 28.065   & 36/500 \\
				\hline 
				
			\end{tabular}
		\end{center}
        		\caption[Differences between Markov model AIC scores and log-linear AIC scores after an accelerated forward model selection for simulated data.]{
                The first column describes how many two-way interactions were included in the Markov model generating the data. $N$ denotes the true population size.
                Columns 3-5 describe the following respective properties of the set of differences of estimates of total population size obtained between fitting the Markov model and log-linear model using an accelerated forward selection based on AIC to the simulated data ($\Delta\hat{N}$): its median, its first to third quartiles interval, the 95$\%$ quantile of its norm. 
                The last column denotes how many times different models are selected between the methods out of the 500 replicates.
                }
		\label{diff_estimates_ms_acc}
	\end{table}

\subsection{Absorbing list} \label{subsection_simabso}

\begin{figure}[htb!]
    \centering
    \includegraphics[scale = 0.63]{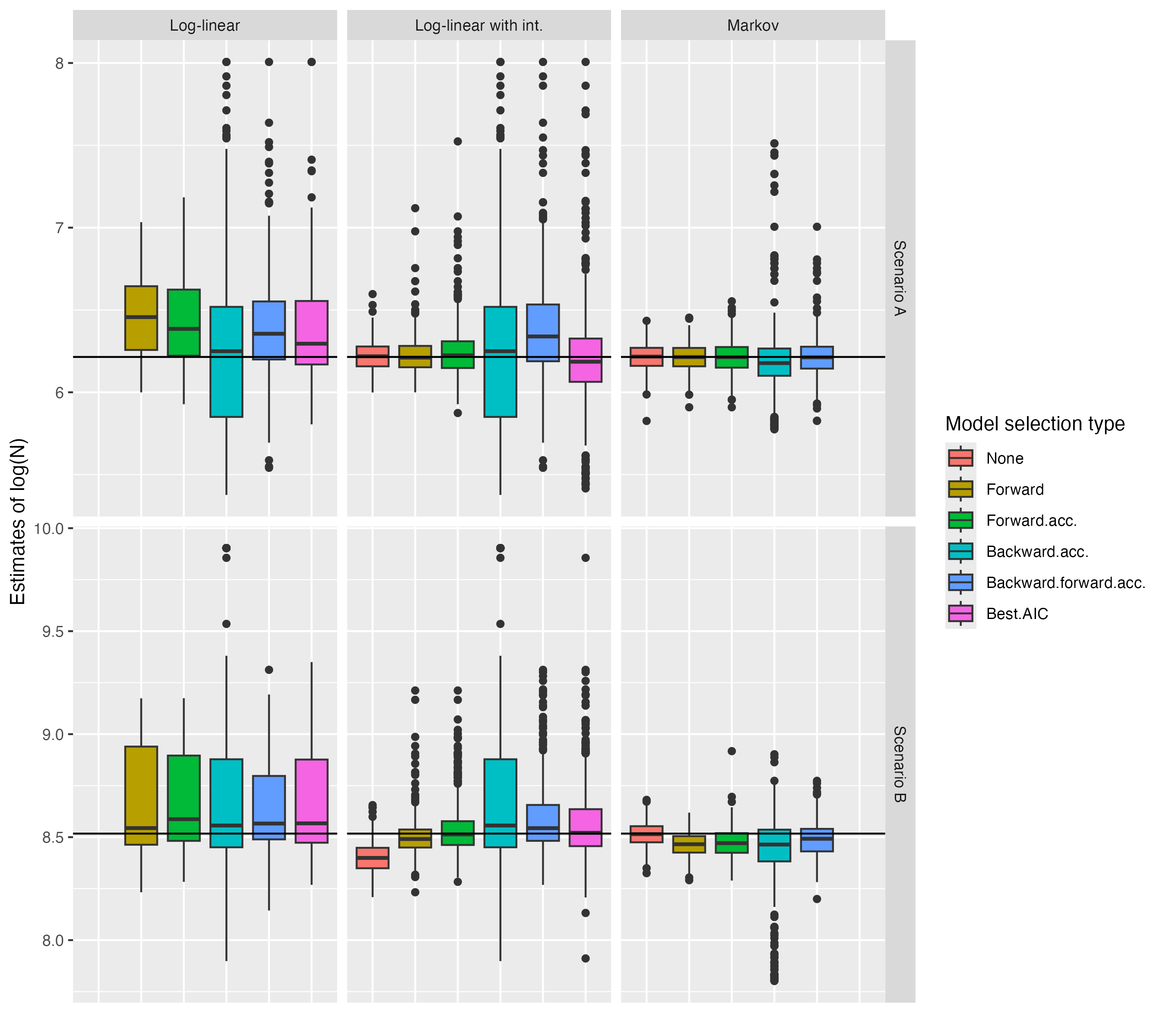}
    \caption{Estimates of the log of the total population size $N$ for 500 simulated data sets from scenarios A and B, using the log-linear model, the log-linear model with forced two-way absorbing interactions and the continuous-time Markov chain model for an absorbing list. The black lines denote the true values of the log of parameter $N$.}
    \label{figure_simulation_sco1}
\end{figure}

		We present simulation results on the analysis of multi-list data with an absorbing list.
        We chose several four-list scenarios of the continuous-time Markov chain model for an absorbing list and we present results for two of them in Figure \ref{figure_simulation_sco1}. For the description of all the scenarios studied and their associated results, see Tables \ref{sim_scenarios} and \ref{sim_expectedcells} and Figures \ref{abso_sco2} to \ref{abso_sco7} in Appendix \ref{appendix_absorbing_sim}. 
       
        For each scenario, we simulated 500 data sets and compared MLEs of the total population size obtained by fitting the Markov model with an absorbing list and the log-linear model. We compared several model selection approaches based on AIC in both cases: (i) choosing directly the true model with the interactions corresponding to those of the generative model; (ii) using a standard stepwise forward selection; (iii) using an accelerated stepwise forward selection; (iv) using an accelerated stepwise backward selection; (v) using an accelerated stepwise backward forward selection, starting with a stepwise backward selection from the model containing all two-way interactions until the backward algorithm stopped and then performing a stepwise forward selection to add potential three-way interactions; (vi) for the log-linear case only using a selection of the model with the best AIC score. In the selection of a Markov model, we did not consider the comparison between all models as the computational cost was prohibitive. 
		
		For the log-linear approach, we also modified the model selection by forcing all two-way interactions containing the absorbing list in the selected model. We will call such interactions \textit{absorbing}. In the case of a forward selection, instead of starting the stepwise algorithm with the null model, we started with the model containing exactly all two-way absorbing interactions. In the case of a backward selection, we did not allow steps that would erase such an interaction. When comparing all models, we simply restricted the comparison to all models containing all two-way absorbing interactions.

		We notice a strong tendency of the log-linear method to overestimate the total population size, with variance of the estimates usually much larger in that case than for the other methods. Furthermore, the choice of the log-linear model with the best AIC does not generally give better results than the other log-linear model selections. We can deduce that the AIC score is not a good indicator of relative fit of the log-linear model when applied to data with a structure similar to that described by the Markov model with an absorbing list (the BIC score was also tested and did not lead to better results). This conclusion further motivates the idea that when an absorbing list is present in the data, it should be specifically modelled, as not accounting for it could lead to significantly biased estimates. 
		
		Results from the Markov chain model have the least variability amongst the methods. The model may tend to slightly underestimate the total population size, where the forward stepwise selection shows the most underestimation, which is lessened by an accelerated forward selection or a bidirectional selection. In those cases, backward selection does not give better results and has larger variability. 
		
		The log-linear model with forced interactions generally gives better results than those obtained when all models are considered. The forward and accelerated forward selections give quite precise estimates, often close to the same selections as the Markov chain model. Their variances and particularly the number and scale of the outliers are logically larger than those of the Markov model, as those log-linear models will most often contain more parameters. Interestingly, the chosen model containing the absorbing two-way interactions as well as the interaction parameters of the generative Markov model often shows an underestimation of the total population and we can conclude that it is not the best log-linear model to fit to the data.

		We conclude from the simulations that taking into account the absorbing list is necessary if the data has a similar structure to that described by the presented Markov chain model with an absorbing list. An accelerated forward selection or an accelerated backward forward model selection should be favoured for the Markov model. The log-linear model with forced absorbing two-way interactions with an accelerated forward selection is a good alternative to the Markov model to account for the absorbing list.

\section{Applications} \label{section_applications}

We present two applications of the continuous-time Markov chain model on multi-list data. In Section \ref{subsection_app_abso}, we study the epidemiological dataset introduced in Section \ref{introduction} which contains an absorbing list and in Section \ref{subsection_app_orderd}, we analyse the dataset on drug use with ordered lists. 

\subsection{Absorbing list application: Stroke data, Northwest England 1994-1995} \label{subsection_app_abso}

		We present in Table \ref{firsteverstroke_results} the results from fitting the Markov and log-linear models to the stroke data collected in Northwest England and described in Section \ref{introduction}. We compare estimates of the total number of cases, $N$, by applying the following methods. When obtaining MLEs from the continuous-time Markov chain model for an absorbing list, we used a standard forward AIC model selection and an accelerated forward AIC model selection and an accelerated backward forward AIC selection from the model with all two-way interactions. When fitting the log-linear model, we used a forward and accelerated forward selection from AIC 
        as well as a forward AIC and accelerated forward AIC procedures from the model containing all two-way absorbing interactions, in this case that is the model containing interactions $\alpha_{13},\alpha_{23},\alpha_{34}$ and $\alpha_{35}$ as list $L_3$ is absorbing. We computed the standard errors from the Hessian to calculate confidence intervals. 
        
	\begin{landscape}
			\begin{table}
				\begin{center}
					\begin{tabular}{l c c c c}
						\hline
						Method & Model selection & Selected parameters & $\hat{N}$ & $95\%$ CI \\
						\hline
						MC & For. AIC & $14,23,24,45$ &  815.57 &  $[738.61, 892.54]$ \\
						MC & Acc. for. AIC &  $12, 14, 15, 23, 24, 25, 45, 124$ & 777.39 & $[671.96, 874.82]$ \\
						MC & Acc. back. for. AIC &  $12, 13, 14, 15, 23, 24, 25, 45, 124$ & 793.01 & $[675.66, 910.37]$ \\
						LL & For. AIC & $12, 14, 23, 24, 25, 34,45, 124, 245$ & 1015.78 & $[892.45, 1199.84]$\\
						LL & Acc. for AIC & $12,13,14,23,24,25,34,35,45,$& 1000.41 & $[853.17, 1250.31]$\\
						& & $124,134,245,345$ & \\
						LL & For. AIC with forced int. & $12,13,14,23,24,25,34,35,45$ & 997.15 & $[851.5, 1244.08]$ \\
						& & $124,134,245$ \\
						LL & Acc. for. AIC with forced int. &  $12,13,14,23,24,25,34,35,45$ & 1000.41 & $[853.17, 1250.31]$ \\
						& & $124,134,245,345$ \\
						\hline
					\end{tabular}
				\end{center}
                	\caption[Results of the analysis on strokes data.]{Analysis of the data of first ever stroke in the Northwest of England July 1994 -June 1995. Here, we write $ij$ or $ijl$ for interaction parameters between lists $L_i$ and $L_j$ or $L_i$,$L_j$ and $L_l$ respectively.}
				\label{firsteverstroke_results}
			\end{table}
		\end{landscape}

		We see that the Markov chain method gives slightly smaller estimates for the total population size, in line with the results of the simulation study. However, the log-linear forward selections from the null start or with absorbing two-way interactions do not differ significantly and almost all models result in overlapping confidence intervals of population size estimates. The log-linear methods select more interaction parameters than the Markov methods.

\subsection{Ordered list application: Drug use in the City of London 2018-2019} \label{subsection_app_orderd}

		We analyse the dataset on drug use in the City of London. Recall that the data is formed of two lists: criminal justice system and community treatment. Separating the community treatment into two ordered lists as described in Section \ref{subsection_account_ordered}, we obtain the three-list dataset presented in Table \ref{drugabuse_data_cityoflondon}. There are two zeros in the table. These are not observed data points as it is impossible for individuals to contribute to those list combinations due to the structure of the data collection. 
		We therefore have only six non-trivial cells in the contingency table and this will limit the number of parameters we can estimate from the data.

		As introduced in Section \ref{subsection_account_ordered} and illustrated in Figure  \ref{drugabuse_graph1}, there are five transition rates and interaction parameters to estimate as well as the total size of the population, $N$. Hence we have a total of six parameters describing six contingency table cells. As choosing to automatically not include one of the two interaction terms in the model would be arbitrary, we decide to simply stop the model selection process after the first parameter is selected.
		
		We compared results obtained using this Markov chain model with results from the log-linear model. We applied both a naive log-linear model with three lists with a one step model selection as in that case it is only possible to add one two-way interaction, and a log-linear model with two lists and a covariate parameter $c_2$ for individuals leaving treatment. In that case we compare the three following models: the model without the covariate parameter and without the interaction between lists, the model with only one of those parameters and the model with both. Results are displayed in Table \ref{drugabuse_results}.

		\begin{table}
			\begin{center}
				\begin{tabular}{l c c c c c}
					\hline
					Method  & Parameters & AIC   & $\hat{N}$ & $95\%$ CI \\
					\hline
					MC & none &  32.89  & 340.53 & $[198.54, 482.53]$\\
					MC & $12$ & 34.75  & 404.76 & $[0, 1196.89]$\\
					MC & $13$ & 30.83  & \textbf{421.22} & $[190.97, 651.47]$\\
					\hline
					LL 3 lists & none &  113.54  & 248.53 & $[113.54, 322.0]$\\
					LL 3 lists & $12$ & 102.00 & 177.02 & $[177.02, 3.376 \times 10^{20	}]$\\
					LL 3 lists & $13$ & 115.5  & 244.38 & $[204.33, 343.07]$\\
					LL 3 lists & $23$ & 34.89 & \textbf{344.99} & $[250.02,563.52]$ \\
					\hline
					LL 2 lists & none & 31.98 & \textbf{345.00} & $[250.02,563.53]$ \\
					LL 2 lists & $12$ & 33.98  & 196.85 & $[177,Inf]$ \\
					LL 2 lists &  $c_2$ & 32.88  & 345.00 & $[250.02, 563,53]$   \\
					LL 2 lists &      $12, c_2$  & 34.89  & 195.67 & $[177,Inf]$ \\
					\hline
				\end{tabular}
			\end{center}
            	\caption[Results on data on drug use in the City of London.]{Analysis of the data on opiate and/or crack cocaine users with the Markov chain model, log-linear model with three lists and log-linear model with two lists. }
			\label{drugabuse_results}
		\end{table}
		
		The Markov model with interaction between $L_1$ and $L_2$ is selected and has the lowest AIC score overall. The log-linear model with three lists selects very clearly the interaction between $L_2$ and $L_3$. The log-linear model with two lists selects the null model, which gives the same estimate and confidence interval as the model with the covariate interaction. The log-linear model with three lists and interaction between $L_1$ and $L_2$ and both two-list log-linear models containing interaction between $L_1$ and $L_2$ are unusable as they have infinite confidence bounds. The estimate given by the Markov model is larger than that obtained from the log-linear models, however the confidence interval given by the continuous-time Markov model contains the estimates from fitting the log-linear models. 
		
		The rate of OCU per 1000 inhabitants estimated by \citet{djennadjrsssa2024} was 59.74. Considering the total population of the City of London in 2019 to be 8,765 \citep{ONS}, the rate proposed by the Markov chain method is slightly lower at 48.06, however the confidence interval contains the value obtained by \citet{djennadjrsssa2024}.

\section{Discussion} \label{section_conclusion}

The results presented in this paper highlight the importance of taking into account the presence of an absorbing list in multi-list data, such as an emigration or death record. 
In that case, we showed that fitting the standard log-linear model leads to an overestimation of the total population size and to more variability of the results. The continuous-time Markov chain model for multi-list data proposed as an alternative to the log-linear model allows for high adaptability and interpretability of the parameters. Fitting the general Markov model or the log-linear model to standard multi-list data seems to lead to similar results, particularly when the number of interactions is low, and the two models are algebraically equivalent when no interactions are present. The Markov model allows us to describe the expected behaviour of an absorbing list and can be modified to describe other relationships between lists. One of the cases presented in this paper is the ordering of lists. Other relationships between lists could be modelled, such as more general asymmetric interactions, for example in the case of referrals between lists as described in \citet{jonesreferral2014}. It is also possible to model lists with different recording time periods. Hence, the Markov model offers many interesting and new prospects for the analysis of multi-list data.

Concerning the modelling of absorbing lists, the overestimation observed when fitting the log-linear model can be explained by the fact that an absorbing list will more likely be modelled through negative two-way interactions. Indeed, if an individual enters an absorbing state at some point in the recording time period of the lists, they will have less chance to be observed on other lists. Some exceptions may apply, for example, in the case of epidemiological data, it is possible that individuals on hospital records are more likely to have also passed away as the individuals with the most serious conditions will more likely have received medical care. This can be modelled with interactions in the Markov model.

When fitting the Markov model to multi-list data with an absorbing list, we advise using an accelerated forward selection or a bidirectional selection (for example an accelerated backward forward selection), preferably in conjunction with bootstrap when it is computationally feasible. Alternatively, we recommend the use of the log-linear model with a forward or accelerated forward selection with forced two-way absorbing interactions. This option is particularly advantageous when the number of lists is greater than five.

Diagnostic testing for violations of model assumptions is a substantial research field within capture-recapture modelling (see for example, \citealt{jeyam2018test, mccrea2017, kenneth1985}). We have explored the potential for the Markov modelling approach presented in this paper to identify absorbing lists when the remit of lists is undisclosed and some premilinary results are presented in Appendix \ref{appendix_detection}. The possibility of using this new modelling framework as the basis for a diagnostic test to detect absorbing lists is an area of ongoing research. 
We additionally plan to further explore the wide applicability of the Markov framework, which has the potential to overcome many of the limitations of current MSE modelling for multi-list data.

\bibliographystyle{apalike}
\bibliography{references.bib}
\newpage
\section*{Appendix}



\appendix

\setcounter{table}{0}
\setcounter{figure}{0}

{\renewcommand{\thetable}{A-\arabic{table}}
\renewcommand{\thefigure}{A-\arabic{figure}}
\section{Continuous-time Markov chain model for MSE}
\subsection{General continuous-time Markov chain model}

For $k = 3$, the matrix $Q$ described in the main text is equal to 

\begin{equation} \label{q-matrix}
  Q = \begin{pmatrix}
		-\sum\star & \lambda_1 & \lambda_2 & \lambda_3 & 0 & 0 & 0 & 0 \\
		0 & -\sum\star & 0 & 0 & \lambda_2\mu_{12} & \lambda_3\mu_{13} & 0 & 0  \\
		0 & 0 & -\sum\star & 0 & \lambda_1 \mu_{12} & 0 & \lambda_3 \mu_{23} & 0 \\
		0 & 0 & 0 & -\sum\star & 0 & \lambda_1\mu_{13} & \lambda_2\mu_{23} & 0 \\
		0 & 0 & 0 & 0 & -\sum\star & 0 & 0 & \lambda_3\mu_{13}\mu_{23}\mu_{123} \\
		0 & 0 & 0 & 0 & 0 & -\sum\star & 0 & \lambda_2\mu_{12}\mu_{23}\mu_{123} \\
		0 & 0 & 0 & 0 & 0 & 0 & -\sum\star & \lambda_1\mu_{12}\mu_{13}\mu_{123} \\
	\end{pmatrix}.  
\end{equation}
Here the elements in the state-space have been ordered as follows: $\emptyset, \{L_1\}, \{L_2\}, \{L_3\}$, $\{L_1, L_2\},$ $\{L_1,L_3\}, \{L_2, L_3\}, \{L_1, L_2, L_3\}$ and the term $- \sum \star$ of each line corresponds to the negative sum of all the other values of the line.

\subsection{Accounting for an absorbing list}
\noindent In the case where $k = 3$ and $L_2$ is an absorbing list, we have the following $Q$-matrix: 
	$$Q^2 = \begin{pmatrix}
		-\sum \star & \lambda_1 & \lambda_2 & \lambda_3 & 0 & 0 & 0 & 0 \\
		0 & -\sum\star & 0 & 0 & \lambda_2\mu_{12} & \lambda_3\mu_{13} & 0 & 0  \\
		0 & 0 & 0 & 0 & 0 & 0 & 0 & 0 \\
		0 & 0 & 0 & -\sum\star & 0 & \lambda_1\mu_{13} & \lambda_2\mu_{23} & 0 \\
		0 & 0 & 0 & 0 & 0 & 0 & 0 & 0\\
		0 & 0 & 0 & 0 & 0 & -\sum\star & 0 & \lambda_2\mu_{12}\mu_{23}\mu_{123} \\
		0 & 0 & 0 & 0 & 0 & 0 & 0 & 0\\
	\end{pmatrix},$$
	where the term $- \sum \star$ of each line corresponds to the negative sum of all the other values of the line.

\section{Comparison of the Markov model and the log-linear model} \label{markov_loglin_equiv}

When the data sources are independent, the continuous-time Markov model is equivalent to the independent log-linear model. Indeed, with simple computations developed below, we see that the following substitution of parameters leads to a one-to-one correspondence between the Markov parameters $(N, \lambda_1, ..., \lambda_k) \in [n, \infty[ \times [0, \infty [^k$, where $n := \sum_{I \in \mathcal{I}} n_I$, and log-linear parameters $(\mu, \alpha_1, ...., \alpha_k) \in \mathbb{R}^{k+1}$. 

	Recall that we have $k$ lists $L_1, \dots, L_k$ with corresponding transition parameters $\lambda_1, \dots, \lambda_k$ and parameter $N \geq 0$ which denote the total population size. Let us denote $p_{i} := 1 - e^{\lambda_{i}}$ for all $1 \leq i \leq k$.
	Let $I \in \mathcal{I}$ be such that for some $0 < l \leq k$,
	we have that $I = \{L_{i_1}, L_{i_2},..., L_{i_l} \}$ and let $L_{i_{l+1}},..., L_{i_k}$ be the $k-l$ remaining lists. Then, recalling that we denote by $p_{\emptyset,I}(1)$ the probability for an individual to be seen by all and only the lists in $I \in \mathcal{L}$ at the end of the recording period, we have by Markov chain theory that (see e.g. \citealt{norris97})
	\begin{align*}
		p_{\emptyset,I}(1) &=  \int_0^1 \int_0^1 \dots \int_0^1 \lambda_{i_1}e^{-\lambda_{i_1} t_1} \dots \lambda_{i_l} e^{-\lambda_{i_l} t_l } e^{-\lambda_{i_{l+1}}} \dots e^{-\lambda_{i_k}} dt_1 \dots dt_l \\
		&= p_{i_1} \dots p_{i_l} (1 - p_{i_{l+1}}) \dots (1 - p_{i_k}).
	\end{align*}
	
	We can describe the log-linear parameters $\mu, \alpha_1, \dots, \alpha_k$ with the following expressions of $p_{i}$s and parameter $N$.
	\begin{align*}
		e^{\mu} &= N(1-p_1) \dots (1 - p_k) \\
		e^{\alpha_i} &= \dfrac{p_i}{(1-p_i)} \qquad \text{for }1 \leq i \leq k.  
	\end{align*}
	Or equivalently with the rate parameters $\lambda_i$s: 
	\begin{align*}
		e^{\mu} &= Ne^{- (\lambda_1 + \dots + \lambda_k)} \\
		e^{\alpha_i} &= e^{\lambda_i} -1 \qquad \text{for }1 \leq i \leq k.  
	\end{align*}
	We see that there is a unique one-to-one correspondence between parameters $(\mu, \alpha_1, ...., \alpha_k) \in \mathbb{R}^{k+1}$ and parameters $(N,p_1,...,p_k) \in [n, \infty [ \times [0,1]^k$, as well as with parameters $(N, \lambda_1, ..., \lambda_k) \in [N_0, \infty[ \times [0, \infty [^k$. 
	
	Then, for any $I \in \mathcal{L}$, the expected value of the corresponding cell ${m^{ll}}_I$ given by the log-linear model will be 
	$${m^{ll}}_I = e^{\mu + \sum_{i,L_i \in I} \alpha_i} = N \prod_{i,L_i \in I}p_i \prod_{j,L_j \notin I} (1-p_j) = N p_{\emptyset, I}(1),$$
	which is equal to the expected value of the corresponding cell given by the Markov chain model. This way, we see that the two parameterisations are equivalent in this case.

\section{Additional simulation results}

\subsection{Comparison between the Markov model and the log-linear model} \label{appendix_comparison_sim}

	\begin{table}[H] 
		\begin{center}
			\begin{tabular}{| c | c c c c | c c c c c c |}
				\hline 
				$N$  & $\lambda_1$ & $\lambda_2$ & $\lambda_3$ & $\lambda_4$ & $\mu_{12}$ & $\mu_{13}$  & $\mu_{14}$ & $\mu_{23}$ & $\mu_{24}$ & $\mu_{34}$\\
				\hline
				500 & 0.11 & 0.22 &  0.36 & 0.51 & 0.8 & 1.2 & 0.7 & 1.5 & 1.3 & 0.75 \\
				\hline
			\end{tabular}
              	\caption[Markov model parameters for simulation scenario.]{Continuous-time Markov chain parameter values for a simulation scenario. $N$ denotes the total population size, $\lambda_1, \lambda_2, \lambda_3, \lambda_4$ the main transition parameters and $\mu_{ij}$ the interaction parameters.}  \label{mc_scenario_unique}
		\end{center}
	\end{table}

\begin{figure}[H]
    \centering
    \includegraphics[scale = 0.6]{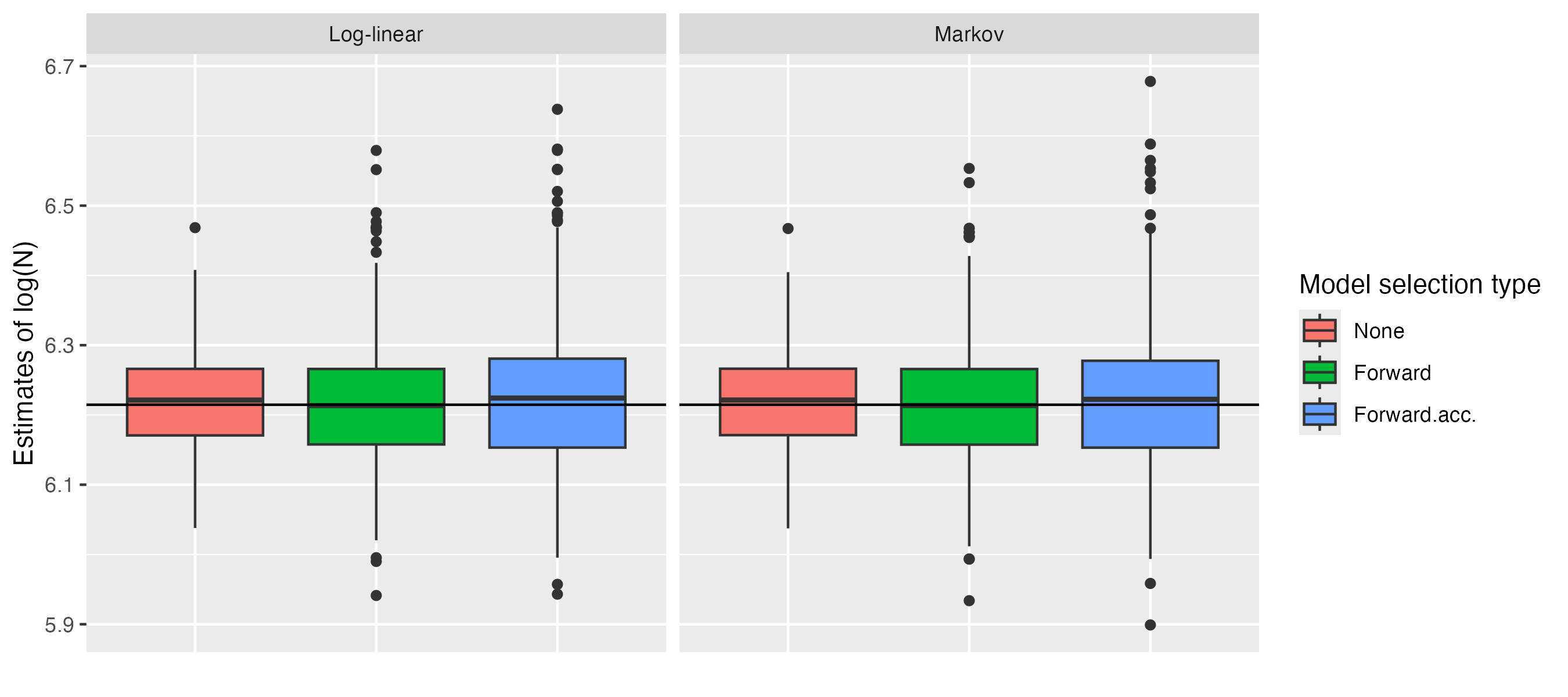}
    \caption{Differences of the log of the total population size estimates between the Markov model and log-linear model fitted to 500 simulated datasets with a forward stepwise model selection based on AIC, when interactions $\mu_{12}, \mu_{13}, \mu_{14}, \mu_{23}$ were included in the generator model described by Table \ref{mc_scenario_unique}.}
    \label{comparison_figure4}
\end{figure}

\begin{figure}[H]
    \centering
    \includegraphics[scale = 0.6]{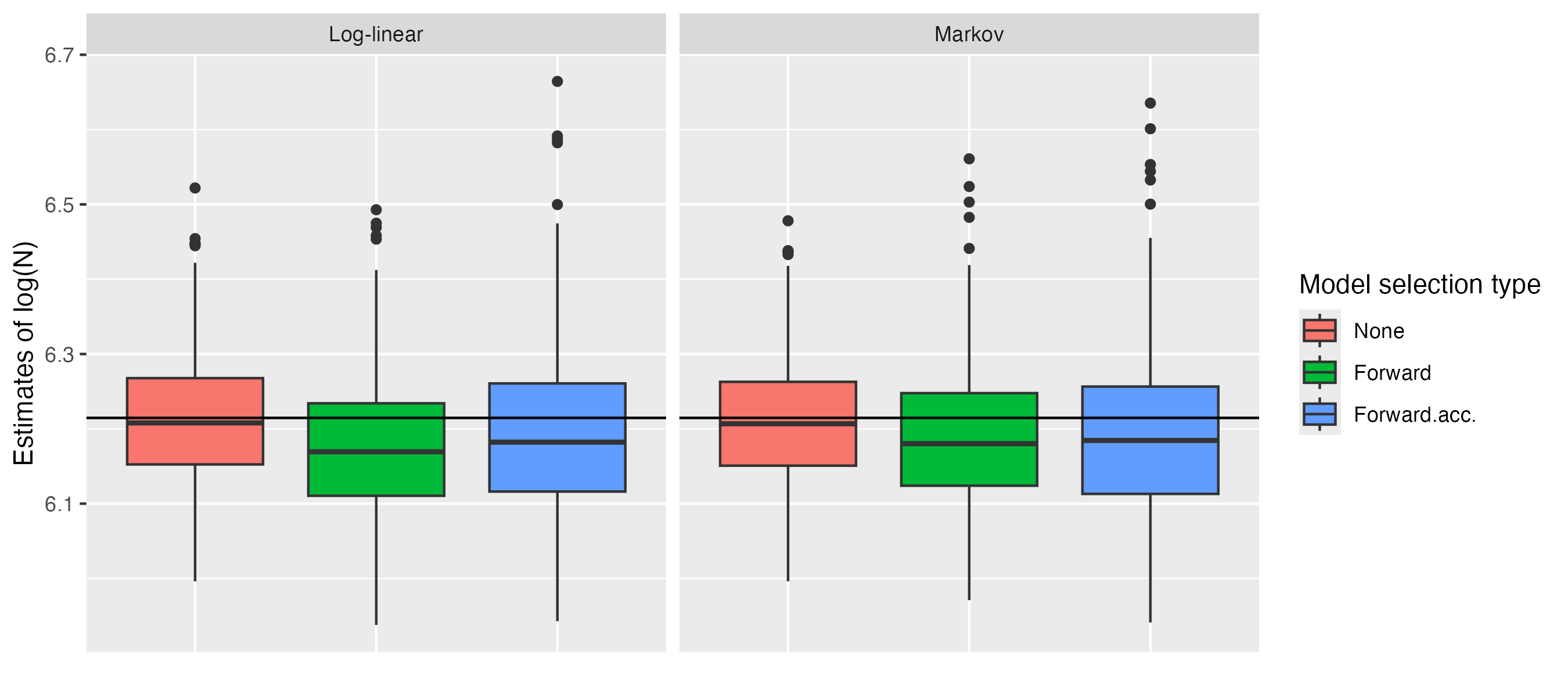}
    \caption{Differences of the log of the total population size estimates between the Markov model and log-linear model fitted to 500 simulated datasets with a forward stepwise model selection based on AIC, when interactions $\mu_{12}, \mu_{13}, \mu_{14}, \mu_{23}, \mu_{24}$ were included in the generator model described by Table \ref{mc_scenario_unique}.}
    \label{comparison_figure5}
\end{figure}

	\begin{table}[H]
		\begin{center}
			\begin{tabular}{| l |  c | c c c c   | }
				\hline 
				$\#$int. & $N$ & med. $\Delta \hat{N}$ &  $[1_{st} Q.,3_{rd} Q.] \Delta \hat{N} $ &  $Q_{0.95} |\Delta\hat{N}|$   & Diff.mo.\\
				\hline
				1 & 500 & 0.01 & [-0.130, 0.180]  & 0.72 & 100 \\
				2 &500 & 0.148 & [-0.129,0.453]  & 1.187 & 99.8 \\
				3 &500 & -0.111 & [-0.245,0.324]  & 0.944 & 100  \\
				4 &500 &  0.381 & [-0.326, 0.488]  & 1.69 & 100 \\
				5 &500 & -0.774 & [-2.560,0.155]  & 7.247 & 89.4\\
				6 &500 & 0.520 & [-2.080,2.672]  & 11.973 & 75.8\\
				\hline 
				
			\end{tabular}
		\end{center}
        		\caption[Differences between Markov model estimates and log-linear estimates for simulated data.]{   The first column describes how many two-way interactions were included in the Markov model generating the data. $N$ denotes the true population size.
                Columns 3-5 describe the following respective properties of the set of differences of estimates of total population size obtained without model selection to the simulated data ($\Delta\hat{N}$): its median, its first to third quartiles interval, the 95$\%$ quantile of its norm. 
                The last column denotes how many times different models are selected between the methods out of the 500 replicates. }
		\label{diff_estimates_noms}
	\end{table}
	
	\begin{table} [H]
		\begin{center}
			\begin{tabular}{| l |  c c c   | }
				\hline 
				$\#$int. & median $\Delta AIC$ &  $[1_{st} Q.,3_{rd} Q.] \Delta AIC $ & $Q_{0.95} |\Delta AIC| $ \\
				\hline
				1 & 0.000 & [0.000,0.000] & 0.000 \\
				2 & 0.000 & [-0.025, 0.027]  & 0.204 \\
				3 & 0.000 & [-0.137, 0.012]  & 0.113 \\
				4 & 0.003 & [-0.026, 0.040]  & 0.192\\
				5 & 0.001 & [-0.045, 0.045]  & 0.212\\
				6 & 0.001 & [-0.055, 0.071]  & 0.255\\
				\hline 
			\end{tabular}
		\end{center}
        	\caption[Differences between Markov model AIC scores and log-linear AIC scores for simulated data.]{Differences between the AIC values obtained from the continuous-time Markov model and the log-linear model without model selection. The first column denotes the number of interactions added in the Markov model to simulate data.}
		\label{diff_AIC_noms}
	\end{table}

	\begin{table} [H]
		\begin{center}
			\begin{tabular}{| l |  c | c c c c   | }
				\hline 
				$\#$int. & $N$ & med. $\Delta \hat{N}$ &  $[1_{st} Q.,3_{rd} Q.] \Delta \hat{N} $ &  $Q_{0.95} |\Delta\hat{N}|$   & Diff.mo.\\
				\hline
				1 & 0.005 & [-0.409, 0.129] & 1.579 & 98.6 & 3/500 \\
				2 & -0.002 & [-0.278, 0.231]  & 2.116 & 97.6 & 6/500 \\
				3 & -0.003 & [-0.264, 0.202]  & 2.026 & 98.2 & 6/500 \\
				4 &  0.001 & [-0.319, 0.214]  & 5.860 & 94 & 11/500	\\
				5 & -0.030 & [-0.648, 0.133]  & 11.335 & 88  & 17/500 \\
				6 &  -0.019 & [-2.212, 1.560]  & 23.287 & 80.8 & 37/500 \\
				\hline 
				
			\end{tabular}
		\end{center}
        	\caption[Differences between Markov model estimates and log-linear estimates after a forward model selection for simulated data.]{The first column describes how many two-way interactions were included in the Markov model generating the data. $N$ denotes the true population size.
                Columns 3-5 describe the following respective properties of the set of differences of estimates of total population size obtained with a stepwise forward model selection based on AIC to the simulated data ($\Delta\hat{N}$): its median, its first to third quartiles interval, the 95$\%$ quantile of its norm. 
                The last column denotes how many times different models are selected between the methods out of the 500 replicates. }
		\label{diff_estimates_ms}
	\end{table}
	
	\begin{table}[H]
		\begin{center}
			\begin{tabular}{| l |  c c c   | }
				\hline 
				$\#$int. & median $\Delta AIC$ &  $[1_{st} Q.,3_{rd} Q.] \Delta AIC $ & $Q_{0.95} |\Delta AIC| $ \\
				\hline
				1 & 0.000 & [0.000, 0.000]  & 0.139 \\
				2 & 0.000 & [0.000, 0.000]   & 0.174\\
				3 & 0.000 & [0.000, 0.000]  & 0.194 \\
				4 &  0.000 & [0.000, 0.000]  & 0.313\\
				5 & 0.605 & [-3.025, 4.292]  & 10.225 \\
				6 & -0.580 & [-4.085, 2.619]  & 10.476\\
				\hline 
				
			\end{tabular}
		\end{center}
        \caption[Differences between Markov model AIC scores and log-linear AIC scores for simulated data.]{Differences between the AIC values obtained from the continuous-time Markov model and the log-linear model with a stepwise forward model selection. The first column denotes the number of interactions added in the Markov model to simulate data.}
		\label{diff_AIC_forward}
	\end{table}

\subsection{Absorbing list simulations} \label{appendix_absorbing_sim}

		\begin{table}[H] 
			\begin{center}
				\begin{tabular}{l | c | c | c c c c c c c c c c c}
					\hline 
					Sce. & $N$ & A.l. & $\lambda_1$ & $\lambda_2$ & $\lambda_3$ & $\lambda_4$ & $\mu_{12}$ & $\mu_{13}$  & $\mu_{14}$ & $\mu_{23}$ & $\mu_{24}$ & $\mu_{34}$\\
					\hline
					1 & 500 & 3 & 0.11 & 0.22 &  0.36 & 0.51 & 1.2 & 1.1 & 1 & 0.9 & 1 & 1 \\
					2 & 1000 & 3 & 0.11 & 0.22 &  0.36 & 0.16 & 1.2 & 1 & 1.4 & 1.6 & 1.2 & 1 \\
					3 & 1000 & 3 & 0.11 & 0.22 &  0.20 & 0.16 & 1.2 & 1 & 1.4 & 1.6 & 1.2 & 1 \\
					4 & 5000 & 2 & 0.11 & 0.22 &  0.20 & 0.16 & 1.2 & 1.13 & 1.4 & 0.9 & 1.15 & 1 \\
					5 & 500 & 3 & 0.11 & 0.22 &  0.21 & 0.16 & 1.2 & 1.1 & 1 & 1.6 & 1 & 1 \\
					6 & 5000 & 3 & 0.12 & 0.36 &  0.11 & 0.22 & 1.1 & 1.13 & 1 & 0.8 & 1.15 & 1.4 \\
					7 & 5000 & 3 & 0.22 & 0.43 & 0.36 & 0.16 & 1.4 & 0.9 & 1.2 & 1.3 & 1.1 & 1 \\
					\hline
				\end{tabular}
			\end{center}
            		\caption[Parameter values for the Markov model with absorbing list scenarios.]{Parameter values for Scenarios 1-7. The absorbing list is shown in the third column (a.l.). For all scenarios, we considered any three way interaction to be equal to 1. Scenarios A and B presented in the main text are scenarios 1 and 4 respectively.}
			\label{sim_scenarios}
		\end{table}
		
		\begin{table}[H] 
			\begin{center}
				\begin{tabular}{l | c c c c c c c c  }
					\hline 
					Scenario & $m_{\emptyset}$ & $m_1$ & $m_2$ & $m_3$ & $m_4$ & $m_{12}$ & $m_{13}$ & $m_{14}$  \\
					
					\hline
					1 & 151.20 & 16.13 & 38.09 & 104.02 & 100.80 &  4.99 &    4.89 &   10.75 \\
					2 & 428.40 & 45.04 &  93.50 & 240.50 & 72.33 & 12.01 &  10.89 &  11.01 \\
					3 & 501.84 &  52.76 & 114.86 & 143.39 &  84.73 &  14.75 &   6.68 & 12.89 \\
					4 & 2509.20 &  260.43 &  806.13 &  553.26 & 426.04 &  44.73 & 65.87 & 64.05\\
					5 & 247.86 & 26.64 & 57.45 & 75.71 & 43.74 & 7.54 & 3.93 & 4.70 \\
					6 & 2242.80 & 272.22 & 949.05 & 362.35 & 533.77 & 128.84 &   18.79 &  64.67 \\
					7 & 1638.00 &  377.73 &  790.89 & 1074.54 &  174.11 & 280.90 &  90.00 & 48.98 \\
					\hline
					
					& $m_{23}$ &  $m_{24}$ & $m_{34}$ & $m_{123}$ & $m_{124}$ & $m_{134}$ & $m_{234}$ & $m_{1234}$\\
					\hline
					1 &     9.04 & 25.39 &  25.08 & 0.78 & 3.33 &   1.80 & 3.36 & 0.35 \\
					2 & 36.44  & 19.39 & 17.21 &   2.87 &   3.63 &   1.61 &   4.57 &   0.61 \\
					3 & 23.04 & 23.82 &  10.57 &  1.85 &  4.46 &  1.00 &  2.95 &  0.40 \\
					4 & 68.14 &   68.60 &  93.87 & 6.16 &  7.75 &  16.22 &  8.32 &  1.25 \\
					5 & 12.24 & 10.14 & 5.73 & 1.10 &  1.33 & 0.42 & 1.33 & 0.14 \\
					6 & 49.88 & 268.71 & 51.12 & 4.20 &  36.47 &  3.82 & 12.13 & 1.19 \\
					7 & 263.50 & 93.62 & 46.54 & 49.29  & 41.00 & 6.98 & 18.84 & 5.09\\
					\hline
				\end{tabular}
			\end{center}
            		\caption[Expected counts for the Markov model with absorbing list scenarios.]{Expected values of contingency table cells for scenarios 1-7. Value $m_{\emptyset}$ indicates the expected number of hidden individuals. Scenarios A and B presented in the main text are scenarios 1 and 4 respectively.}
			\label{sim_expectedcells}
		\end{table}

\begin{figure}[H]
    \centering
    \includegraphics[scale = 0.63]{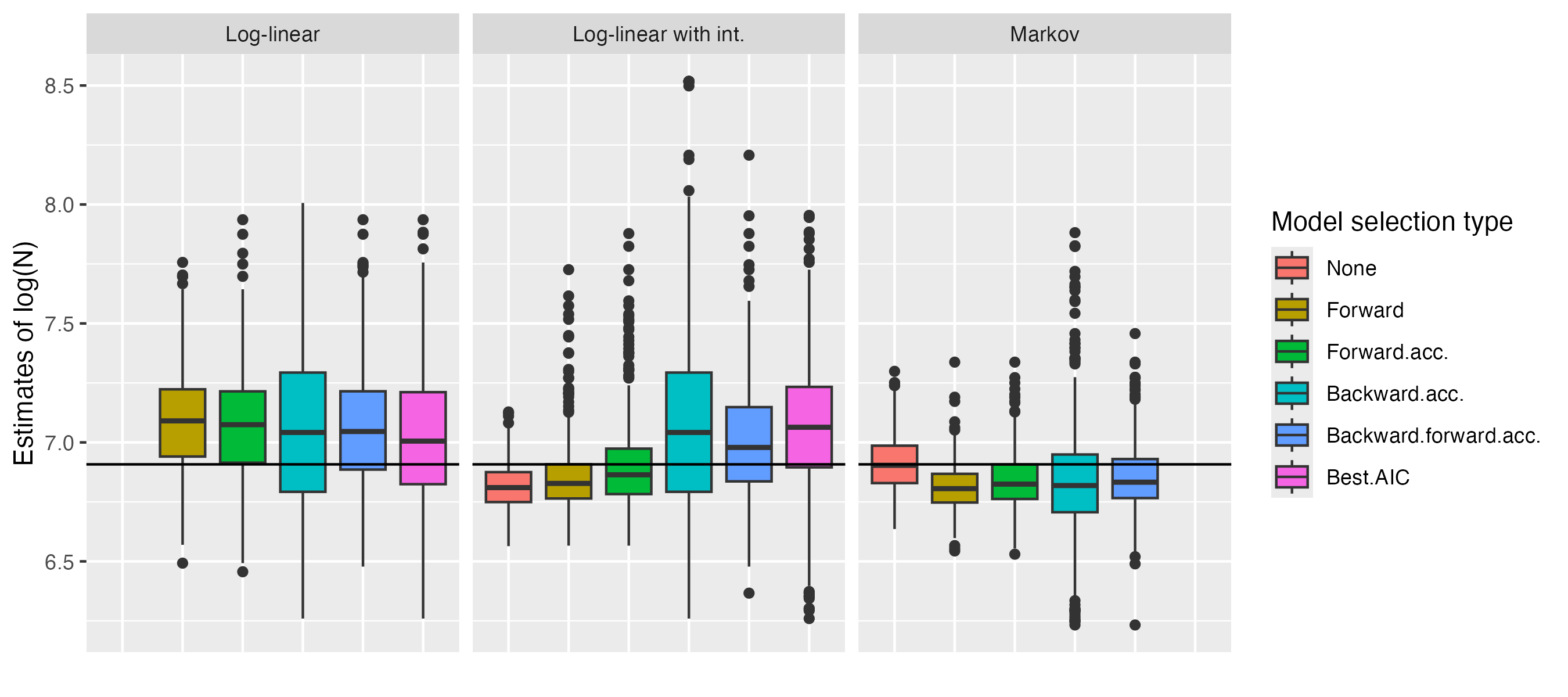}
    \caption{Estimates of the log of the total population size $N$ for 500 simulated data sets from scenario 2, using the log-linear model, the log-linear model with forced two-way absorbing interactions and the continuous-time Markov chain model for an absorbing list. The black lines denote the true value of the log of parameter $N$.}
    \label{abso_sco2}
\end{figure}

\begin{figure}[H]
    \centering
    \includegraphics[scale = 0.63]{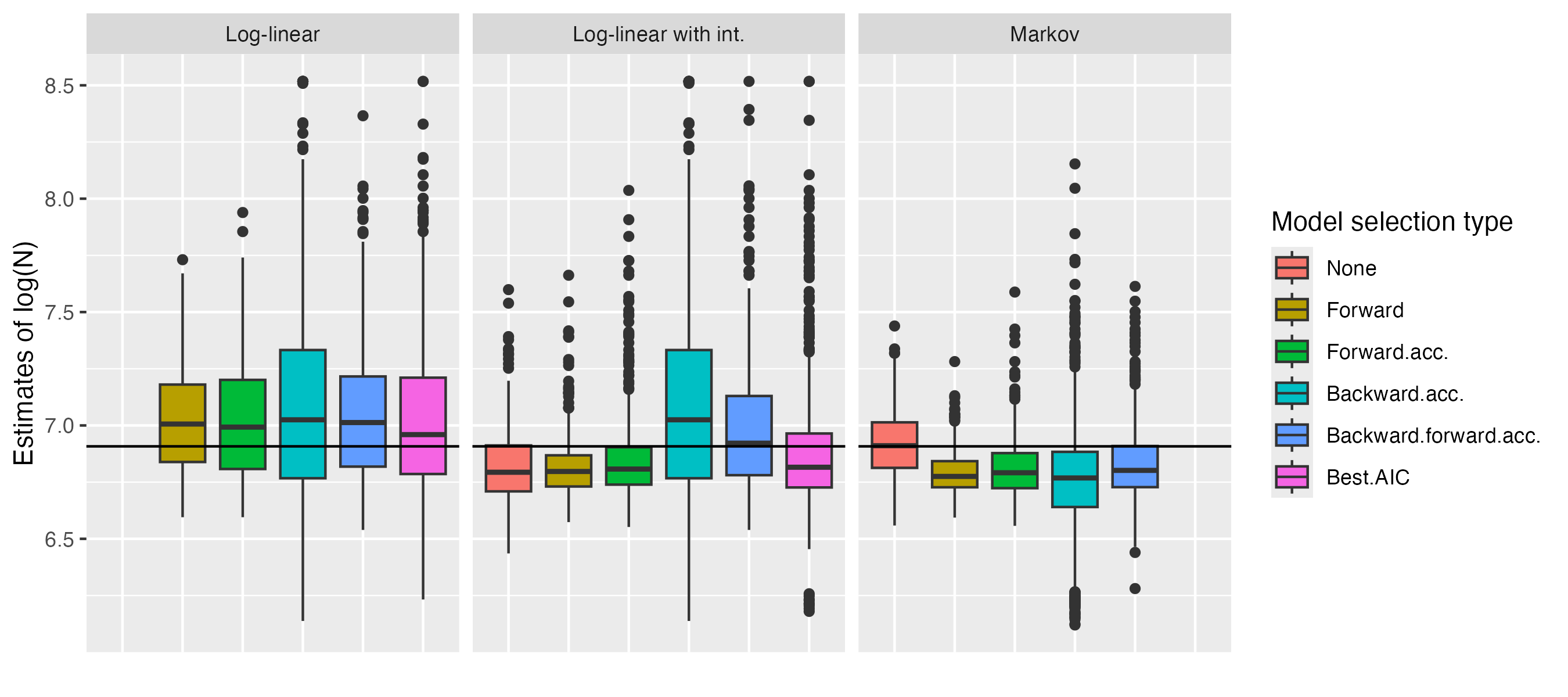}
    \caption{Estimates of the log of the total population size $N$ for 500 simulated data sets from scenario 3, using the log-linear model, the log-linear model with forced two-way absorbing interactions and the continuous-time Markov chain model for an absorbing list. The black lines denote the true value of the log of parameter $N$.}
    \label{abso_sco3}
\end{figure}

\begin{figure}[H]
    \centering
    \includegraphics[scale = 0.63]{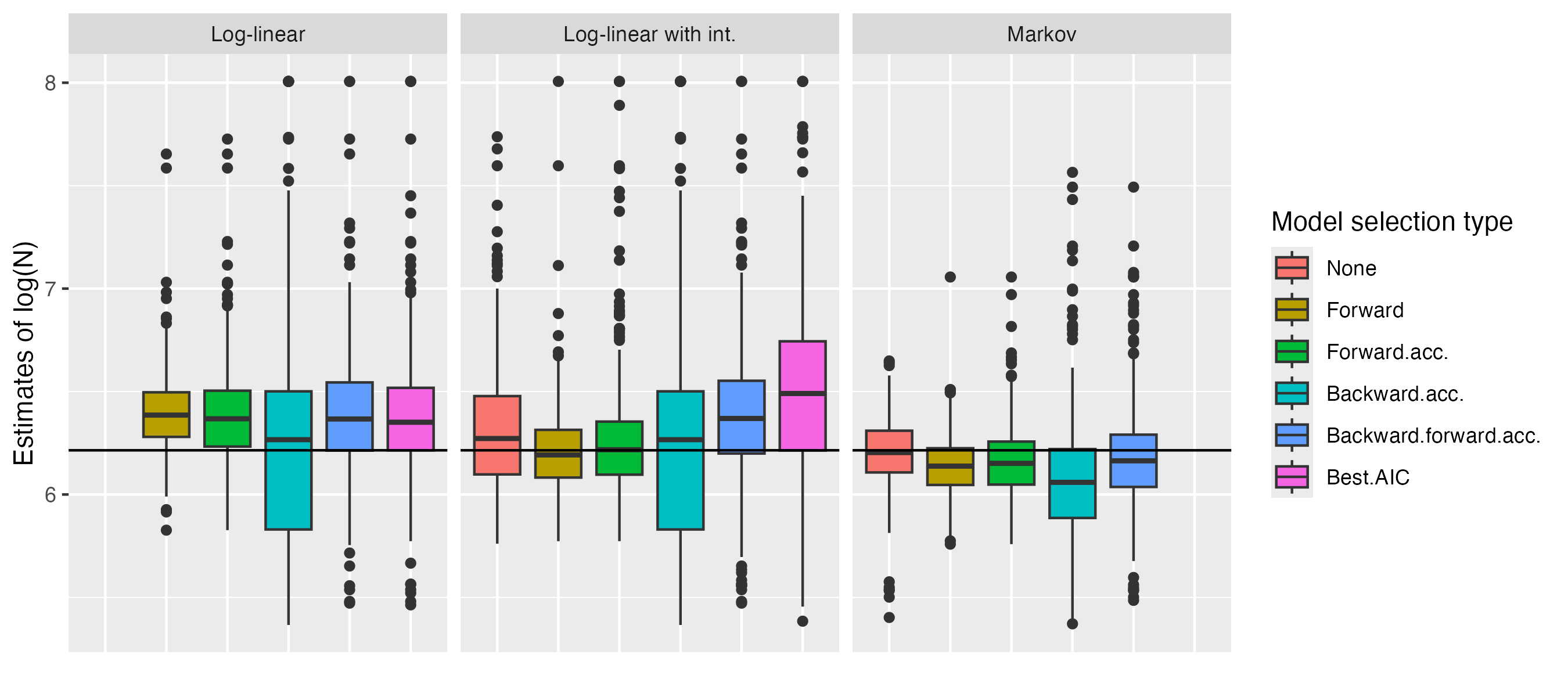}
    \caption{Estimates of the log of the total population size $N$ for 500 simulated data sets from scenario 5, using the log-linear model, the log-linear model with forced two-way absorbing interactions and the continuous-time Markov chain model for an absorbing list. The black lines denote the true value of the log of parameter $N$.}
    \label{abso_sco5}
\end{figure}

\begin{figure}[H]
    \centering
    \includegraphics[scale = 0.63]{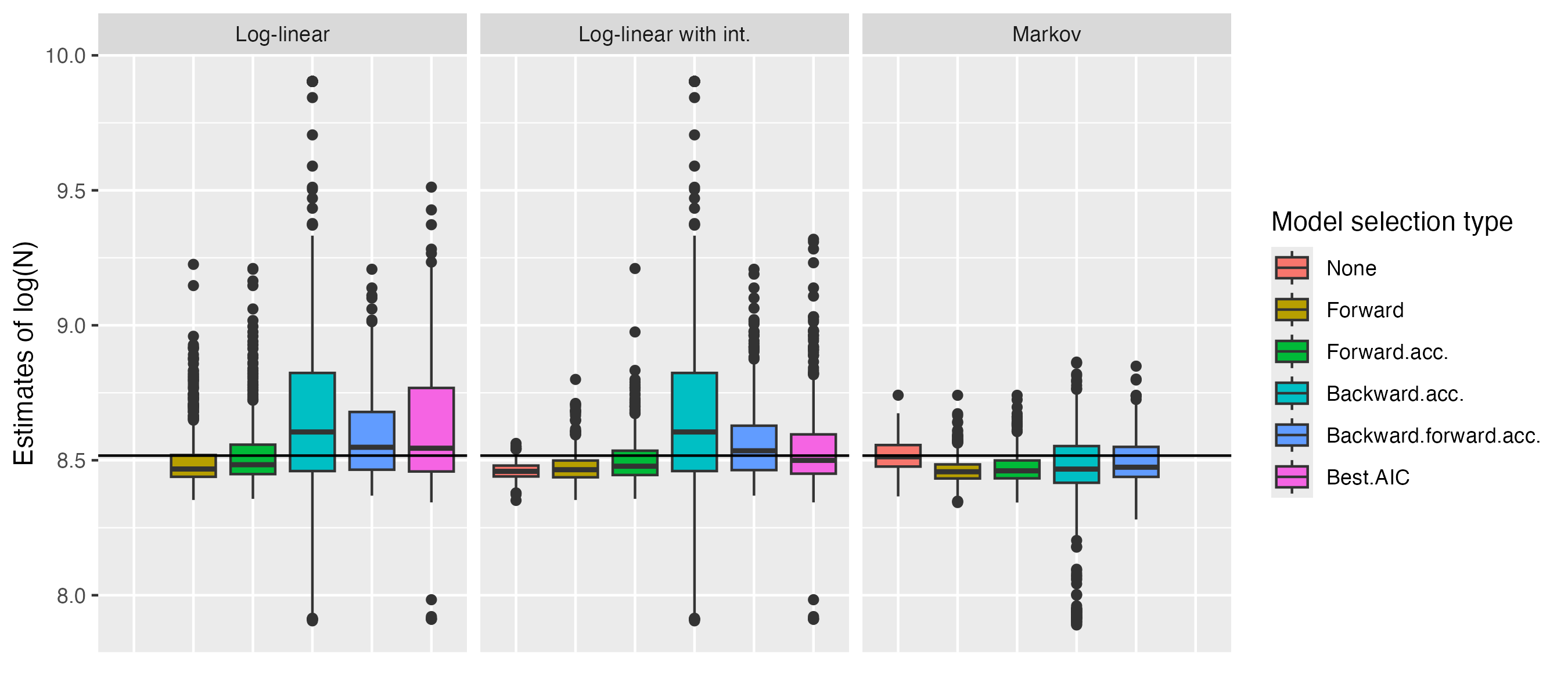}
    \caption{Estimates of the log of the total population size $N$ for 500 simulated data sets from scenario 6, using the log-linear model, the log-linear model with forced two-way absorbing interactions and the continuous-time Markov chain model for an absorbing list. The black lines denote the true value of the log of parameter $N$.}
    \label{abso_sco6}
\end{figure}

\begin{figure}[H]
    \centering
    \includegraphics[scale = 0.63]{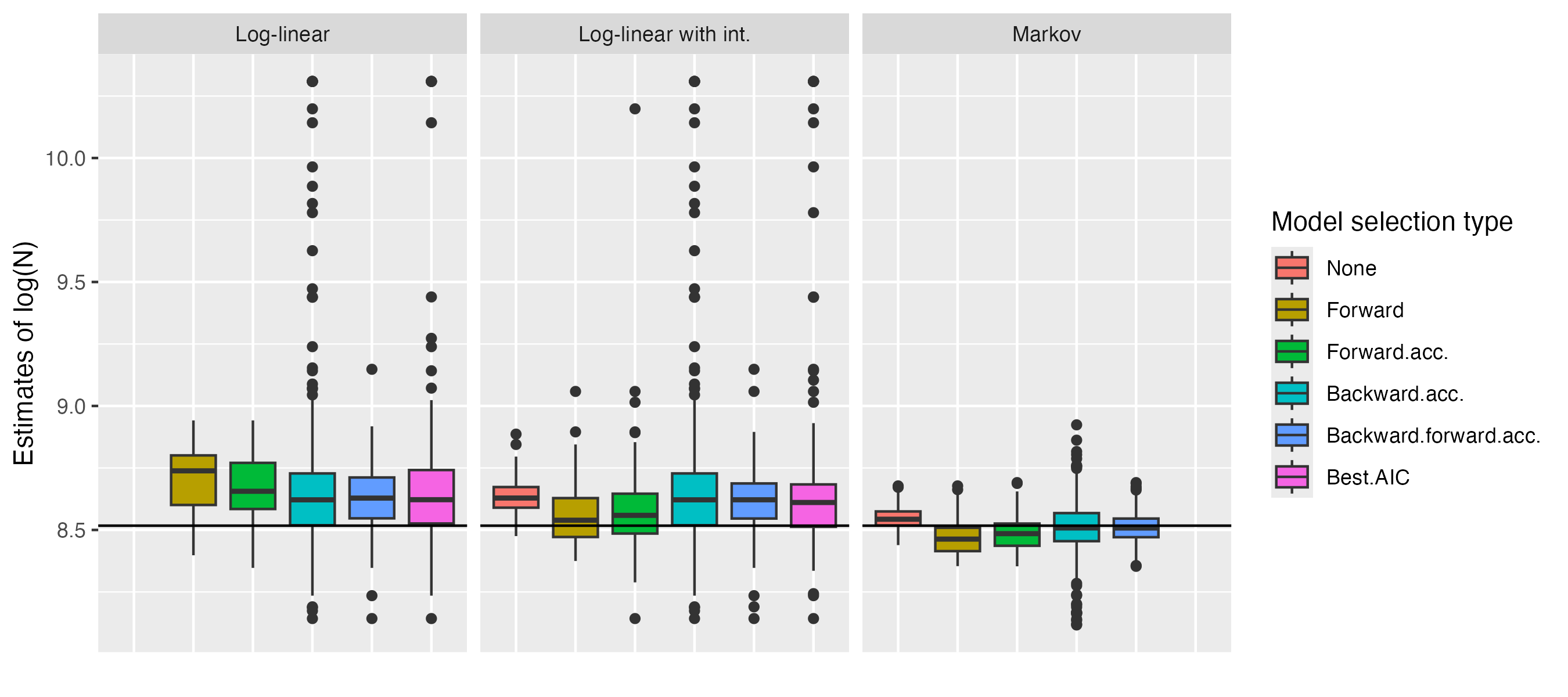}
    \caption{Estimates of the log of the total population size $N$ for 500 simulated data sets from scenario 7, using the log-linear model, the log-linear model with forced two-way absorbing interactions and the continuous-time Markov chain model for an absorbing list. The black lines denote the true value of the log of parameter $N$.}
    \label{abso_sco7}
\end{figure}

        \section{Detection of an absorbing list} \label{appendix_detection}

    Data used for MSE are very sensitive and caution is often necessary to diminish deductive disclosure risk. For this reason, the nature of the lists is sometimes omitted on published data (e.g. \citealt{Bales2020}). As argued above, failing to model the specificity of an absorbing list can lead to biased estimates. In a case where no information on the lists is published, this problem is now exacerbated as no absorbing list is known even if they exist. For this reason, we further explored the extent to which the proposed method could be used to detect the presence of an absorbing list.

               For multiple scenarios, we fitted the log-linear model and the Markov model for an absorbing list to 500 simulated data replicates, varying which list we define as absorbing in the case of the Markov model. For each case, we use the accelerated forward selection based on AIC. We compare AIC values given by each method. We present in Table \ref{detect_abso_table_percentage} what percentage of the time each method obtains the lowest score. We compared only the Markov models and the Markov models with the log-linear model. We also observed the mean number of interactions selected for each method. 
               
		The absorbing list is selected a majority of the time in each case (between $67.4 \%$ and $87.9 \%$). In most cases, adding the log-linear model influences only slightly how often the true absorbing list is selected. We observe an exception in scenario A where the probability for the Markov model with the true absorbing list to be selected drops to $11.8 \%$ and the log-linear model is selected the vast majority of the time.
	
\begin{table}[H]
	\begin{center}
		\begin{tabular}{c | c c c c || c c c c c }
			\hline
			Method & $M_{L_1}$ & $M_{L_2}$ & $\bm{M_{L_3}}$ & $M_{L_4}$ & $M_{L_1}$ & $M_{L_2}$ & $\bm{M_{L_3}}$ & $M_{L_4}$ & LL \\
			\hline
			Scenario A & $5 \%$  & $5.2 \%$ & $87.8 \%$ & $2 \%$ & $0.8 \%$ & $0.4 \%$ & $11.8 \%$ & $0.4 \%$ & $86.6 \%$  \\
			Scenario B & $7.8 \%$  & $5.4 \%$ & $79.6 \%$ & $7.2 \%$ & $6.8 \%$ & $4 \%$ & $73.4 \%$ & $6.4 \%$ & $9.4 \%$ \\
			Scenario C & $84 \%$  & $3 \%$ & $80.4 \%$ & $8.2 \%$ & $6.8 \%$ & $2.4 \%$ & $76.8 \%$ & $7.2 \%$ & $9.4 \%$ \\
			Scenario D & $11.6 \%$  & $5.6 \%$ & $67.4 \%$ & $15.4 \%$ & $8 \%$ & $4 \%$ & $57.4 \%$ & $10.4 \%$ & $20.2 \%$ \\
			Scenario E & $7.6 \%$  & $6.2 \%$ & $79.6 \%$ & $6.6 \%$ & $5.8 \%$ & $6 \%$ & $74.6 \%$ & $5.2 \%$ & $8.4 \%$  \\

			\hline 
		\end{tabular}
	\end{center}
    	\caption[Percentage of time each model is selected for detection of an absorbing list.]{Percentage of time each list is selected as absorbing when comparing the AIC of the Markov chain model with different absorbing lists. The right-hand columns present the percentage of time the Markov model with a certain absorbing list is selected or the log-linear model based on AIC. We used an accelerated stepwise forward selection with AIC. }
	\label{detect_abso_table_percentage}
\end{table}
 }


\end{document}